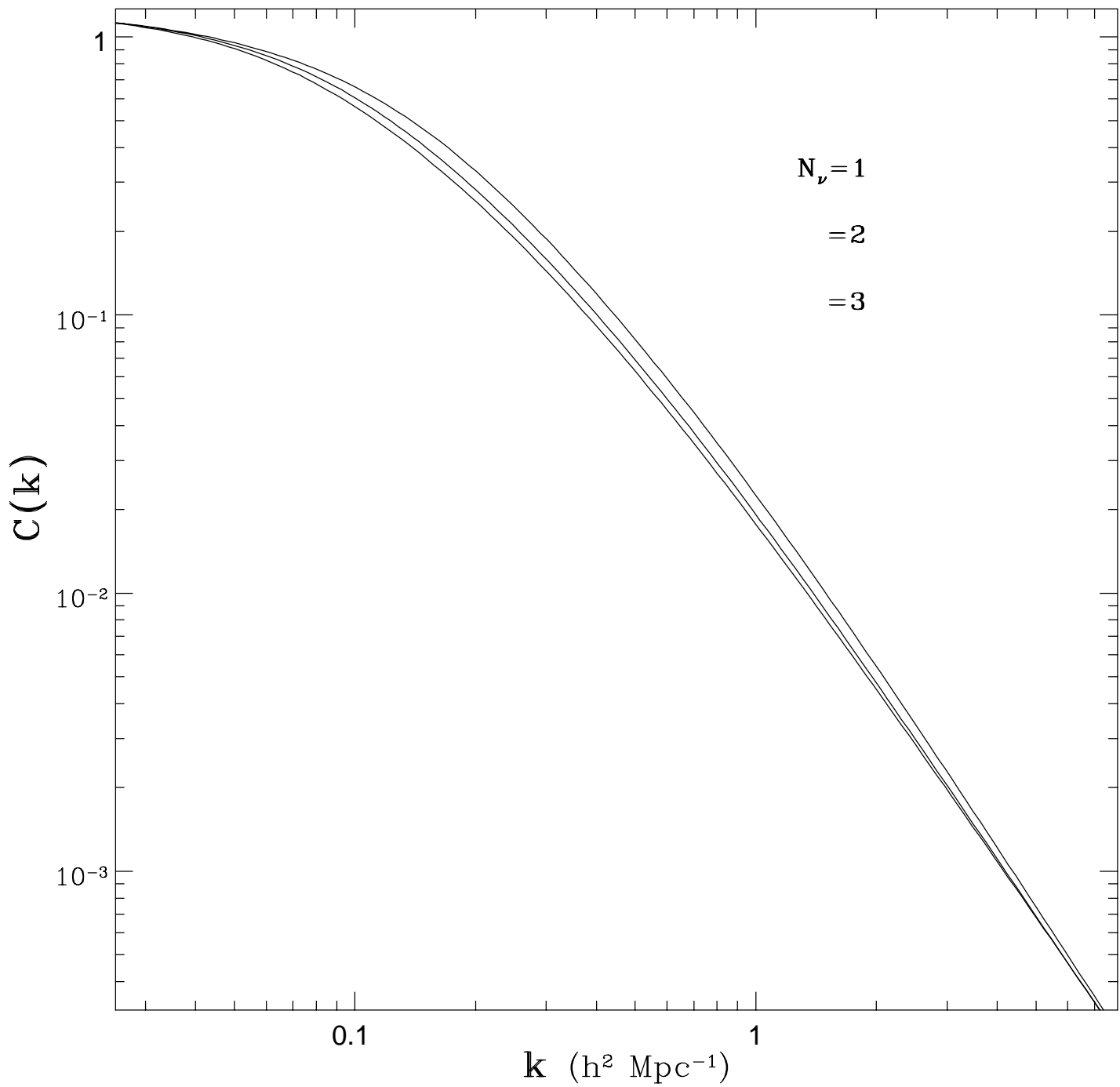

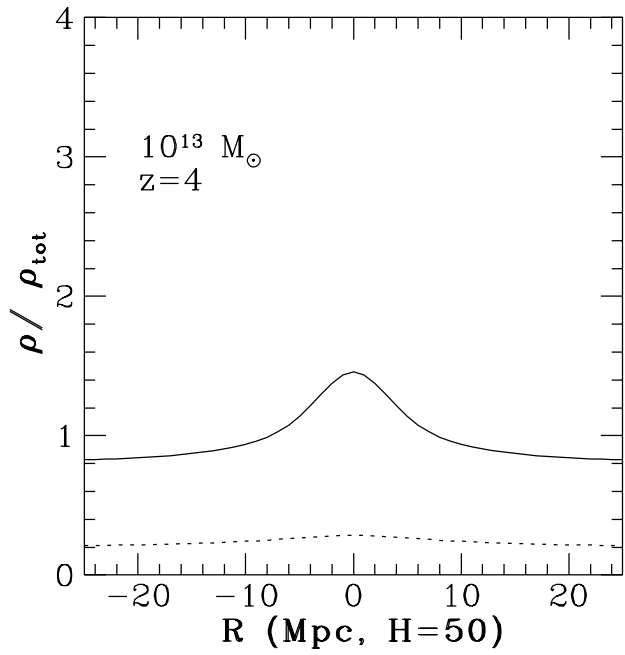
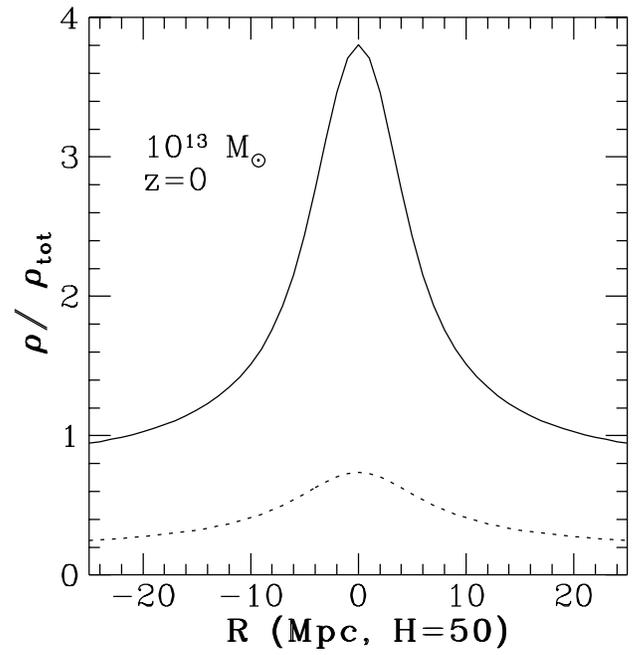
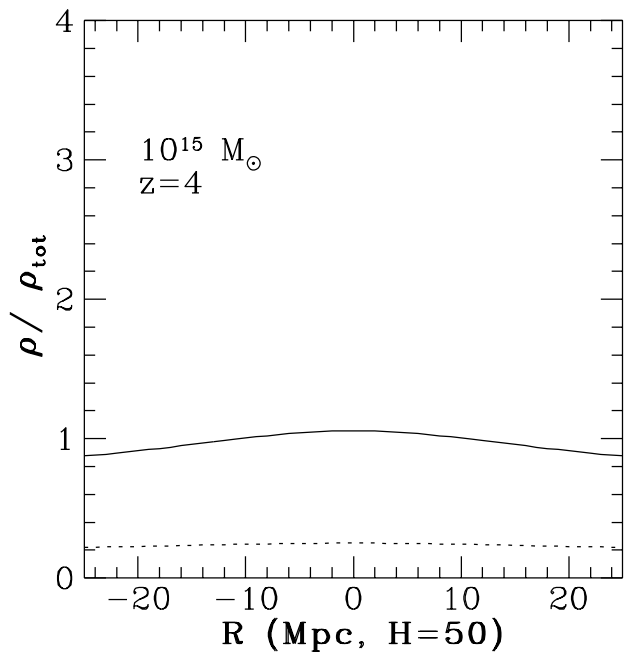
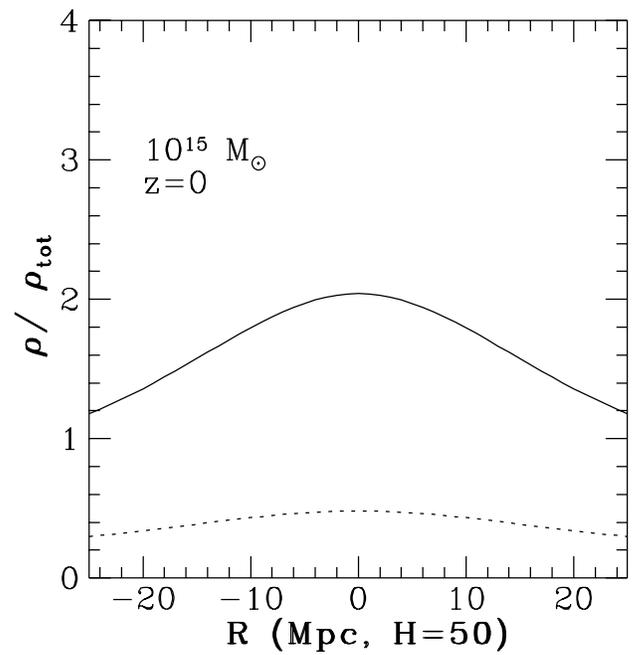

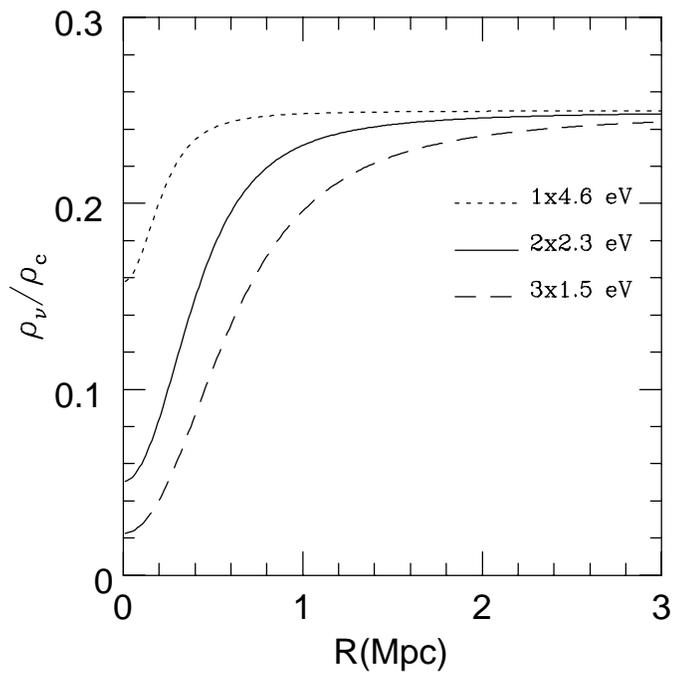 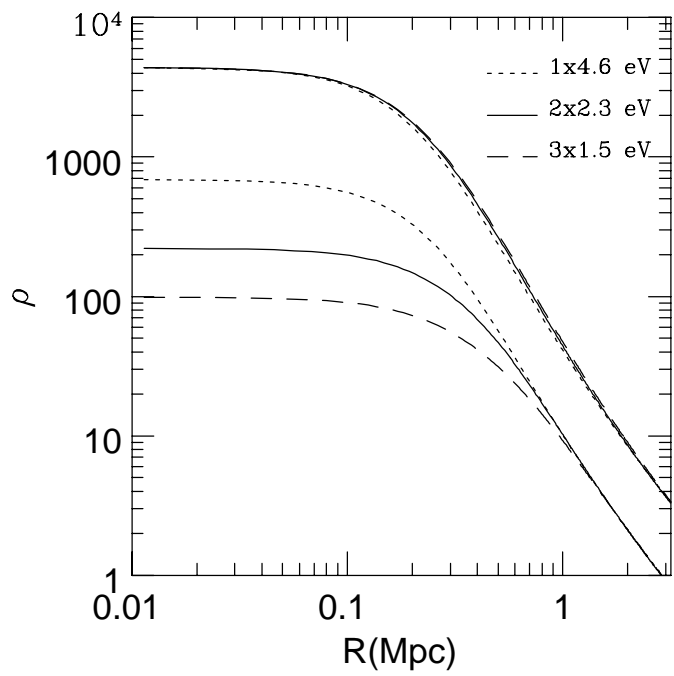

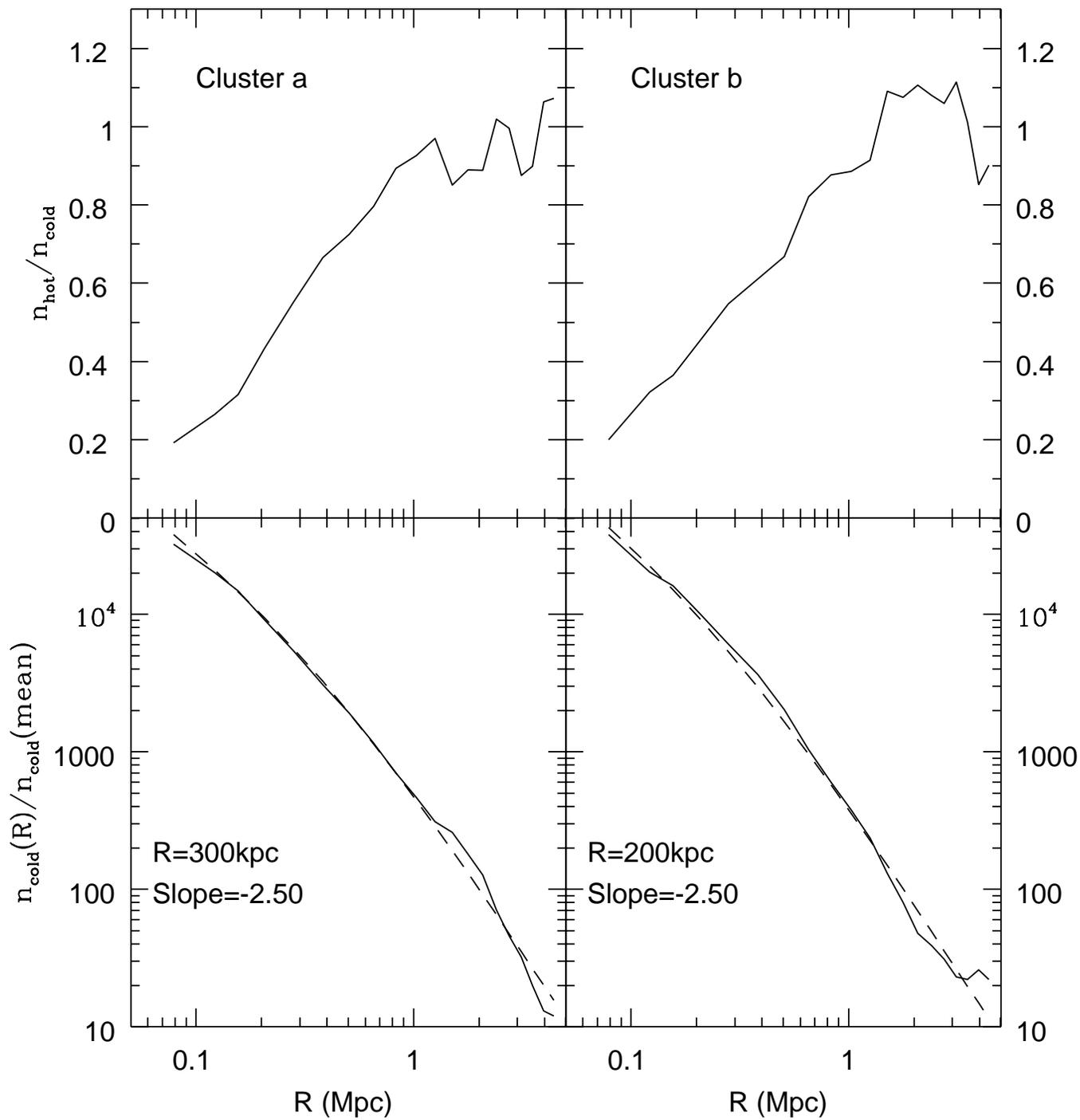

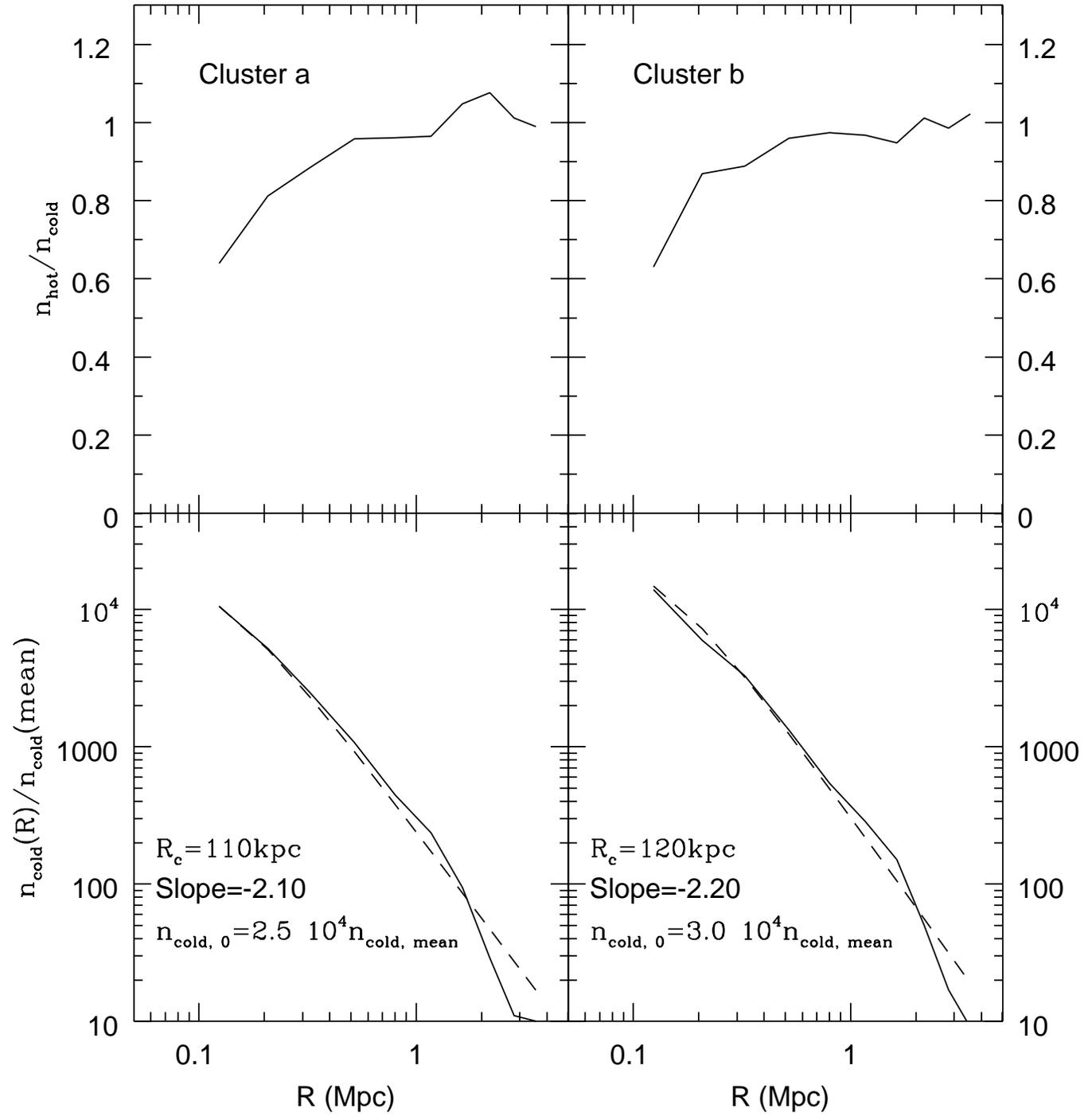

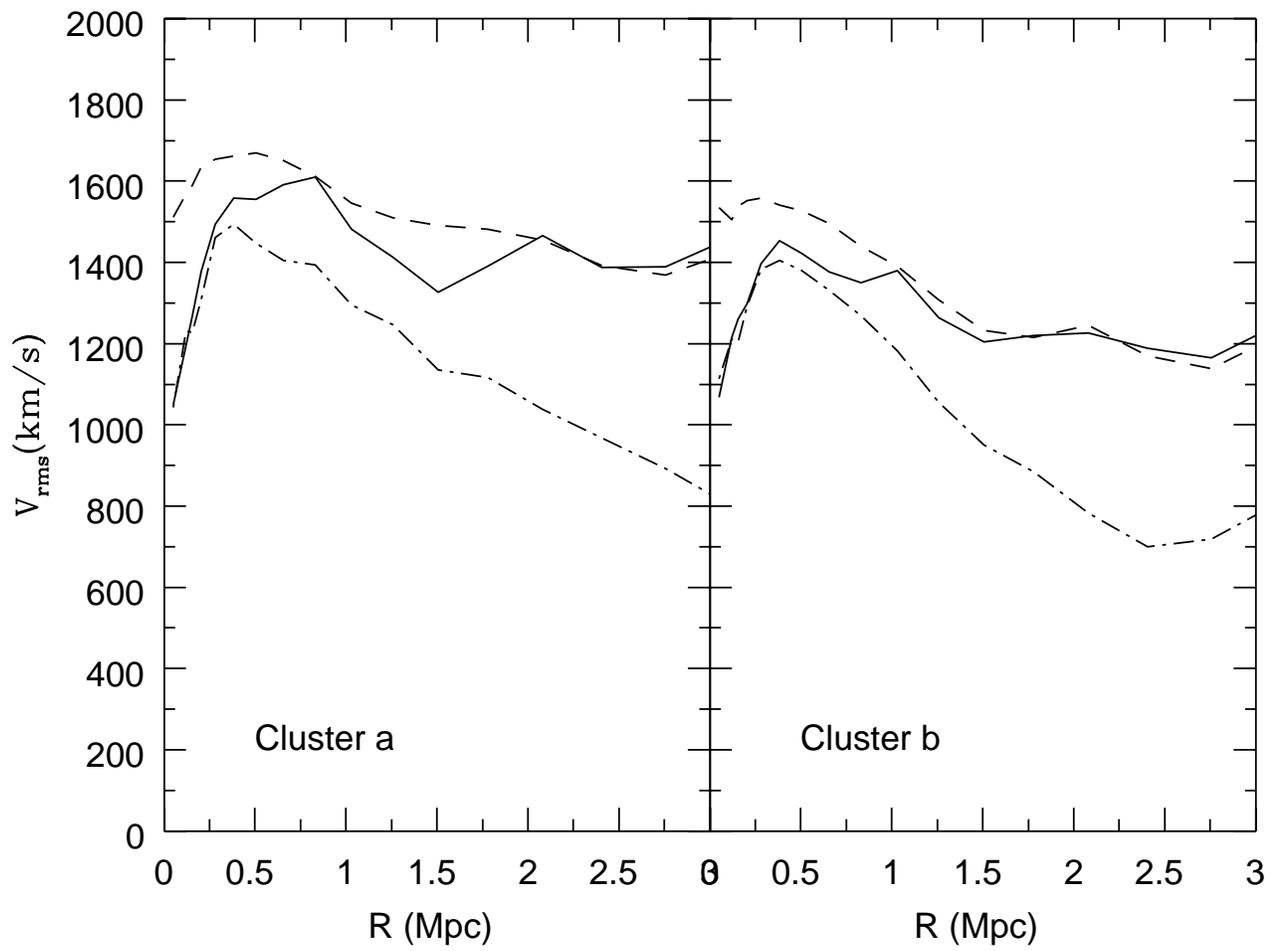

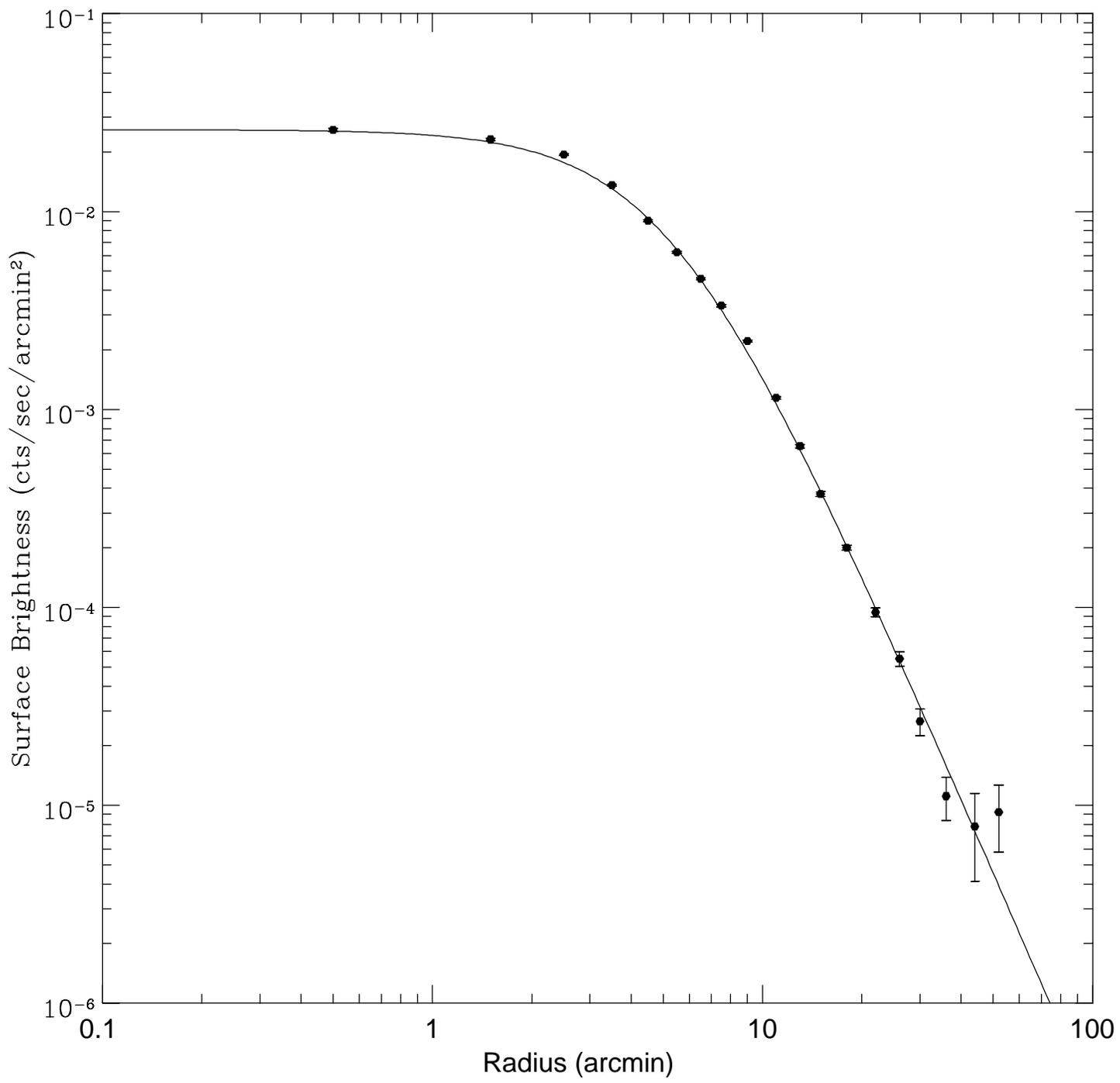

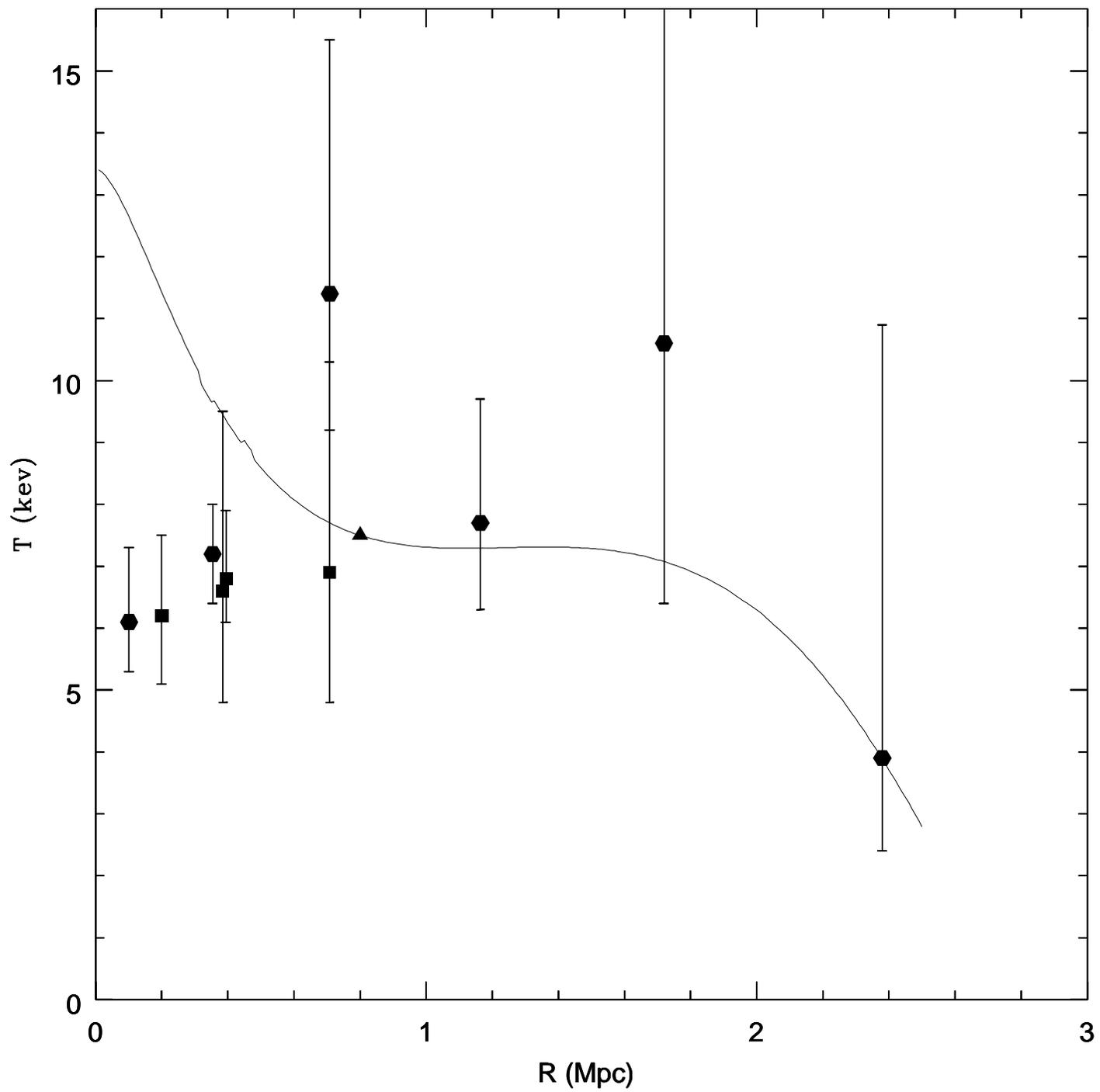

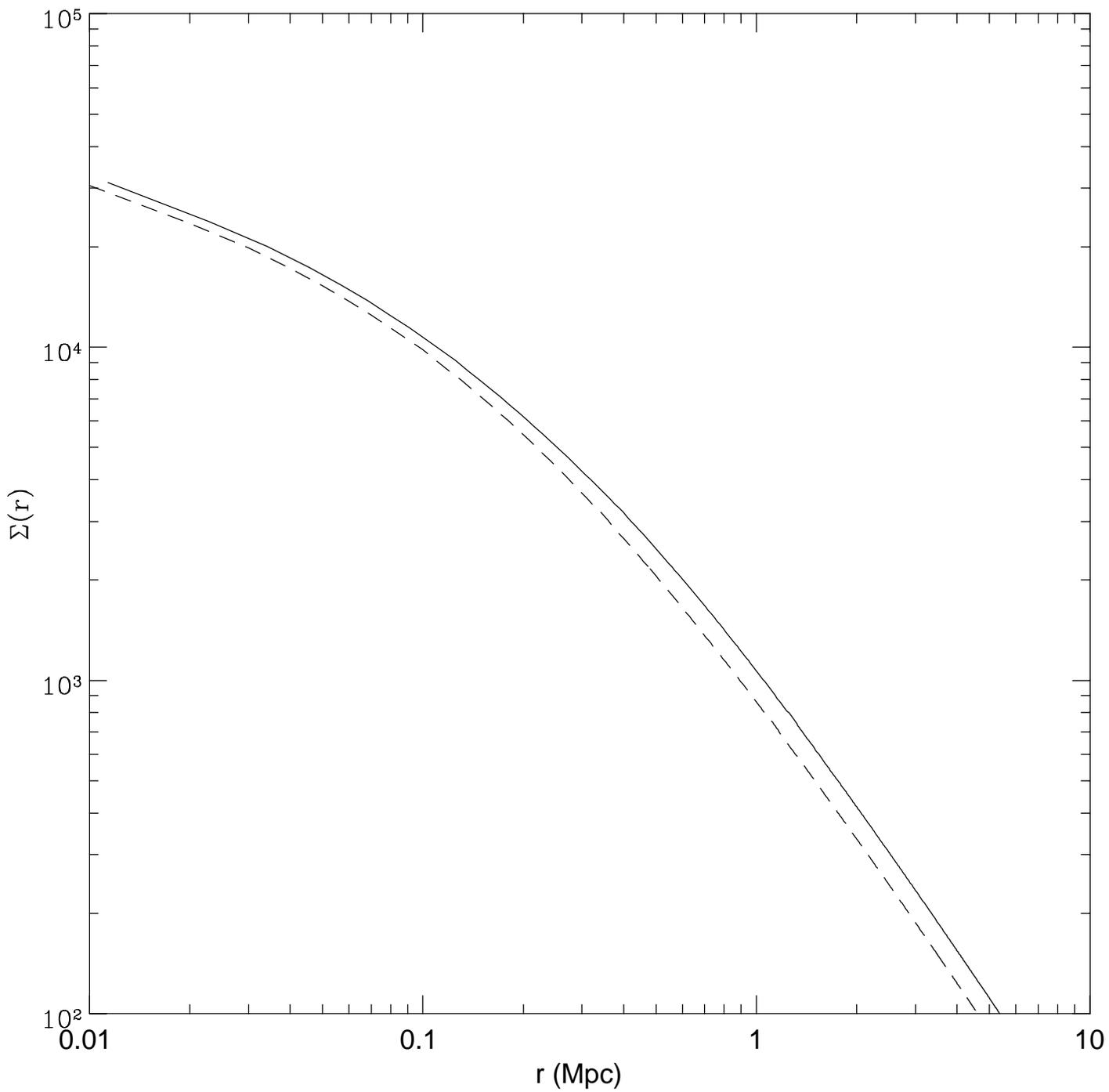

# Mixed Dark Matter in Halos of Clusters

Lev Kofman, [1] Anatoly Klypin, [2] Dmitry Pogosyan, [3] and J. Patrick Henry [1]

[1] Institute for Astronomy, University of Hawaii, 2680 Woodlawn Dr., Honolulu, HI 96822
[2] Astronomy Dept. New Mexico State University, Box 30001, Dept. 4500, Las Cruses, NM88003
[3] Canadian Institute for Theoretical Astrophysics, 60 St George Str., Toronto, M5S 1A7

## ABSTRACT

We discuss the structure of clusters in a class of flat cosmological models with the fraction of mass $\Omega_{CDM} \simeq 0.8$ in cold dark matter, and the rest in hot dark matter in the form of massive neutrinos. These Cold + Hot Dark Matter (CHDM) models with one, two or three massive neutrinos, with total mass $\simeq 4.6$ eV, are considered. Neutrinos of that mass range are too hot and cannot constitute the halos of galaxies and groups, but only of clusters of galaxies. The limit on the density of neutrinos in the central parts of galaxy clusters is estimated from the phase space density constraints. The ratio of the density of neutrinos to that of cold dark matter through the cluster is found analytically. It appears that the density of the neutrinos is suppressed within the Abell radius. However, neutrinos contribute $\sim 20\%$ of the mass density to the cluster halo.

Our numerical simulations fairly match analytical results. The simulations indicate that the cluster halo dark matter density profile has the power-law slope $\alpha \simeq 2.5$, which is close to that in the model with the cosmological constant. We also found that in the CHDM models the velocity dispersion is almost constant across the cluster. This is quite different from the model with the cosmological constant or from the open model where the velocity dispersion in the cluster outskirts declines.

We discuss observational tests that can probe cold, neutrino, and baryonic components unequally distributed in clusters: X-ray emission and weak gravitational lensing. We input the found spherically symmetric fit to the CHDM mass density profile and the X-ray surface brightness for the cluster A2256 into simple model of the hydrostatic equilibrium of the hot gas. The derived theoretical temperature around the center departures from both the observational data and actual prediction of the cosmological model, which give almost constant temperature. We also find that the problem of high baryonic fraction in clusters is not resolved in the CHDM models.

*Subject headings:* theory – dark matter– clusters of galaxies

# 1 Introduction

Most cosmological models are based at present on variations of the Cold Dark Matter scenario, with initial conditions compatible with inflationary theory. Two currently popular models of that type, which satisfy a range of observational data, are the Cold + Hot Dark Matter (CHDM) model and Cold Dark Matter model with a cosmological constant ($\Lambda$CDM). These models have an extra parameter, $1 - \Omega_{CDM}$, which can be adjusted to make the models compatible with the power spectrum of density fluctuations $P(k)$ estimated from galaxy catalogs, large-scale velocity streaming, and CMB $\Delta T/T$ fluctuations. Another set of constraints comes from the abundance of protogalaxies at high redshifts and clusters at the present epoch. The CHDM model assumes a fraction (20%–30%) of the critical density to be in the form of massive neutrinos. In the $\Lambda$CDM model most of the mass of the Universe (60%–80%) is due to the cosmological constant $\Lambda$. In these respects the models are relatively well studied, so that one can put limits on the region of the parameter space in the $(1 - \Omega_{CDM}, h)$ plane, $h$ being the Hubble parameter in units of 100 Mpc km$^{-1}$ s$^{-1}$. An earlier version of the post COBE CHDM model adopted $\Omega_\nu = 0.3$ from one massive (supposedly $\tau$-) neutrino with mass $m_\nu = 7$ eV ( Davis et al. 1992; Klypin et al. 1993). The allowed region of the parameter space $(\Omega_\nu, h)$ was studied by Pogosyan & Starobinsky (1993) and Liddle & Lyth (1993); protogalaxy abundances were studied by Mo & Miralda-Escude (1994); Kauffman & Charlot (1994); Ma & Bertschinger (1994); Klypin et al. (1995), and clusters abundance and their statistical properties by Klypin & Rhee (1994); Jing & Fang (1994); Croft & Efstathiou (1994); Cen & Ostriker (1994); Bryan et al. (1994); Walter & Klypin (1995). At present, the model with values $(\Omega_\nu, h) = (0.2, 0.5)$ has been adopted. A disadvantage of this model is that it is in the conflict with the high value of the Hubble constant (e.g., Freedman et al. 1994).

Some interpretations of current neutrino experiments do not exclude the models of two massive neutrinos with close masses $\sim 2.3$ eV each (Primack et al. 1995). From the general point of view, it is instructive to consider even more exotic combinations with three massive neutrinos. The density parameter of the Universe associated with the massive neutrinos is $\Omega_\nu \approx 0.011 h^{-2} \sum_i (m_{\nu_i}/1\text{eV})$. Because the neutrino Jeans length depends on the neutrino mass (Doroshkevich et al. 1980; Bond & Szalay 1983), the shape of $P(k)$ in the CHDM models depends not only on the total neutrino density $\Omega_\nu$, but also on the masses of neutrinos of different types. Thus for fixed total density $\Omega_\nu$ the spectrum $P(k)$ will depend on how the mass is distributed among three types of neutrinos. Here we will consider the general range of neutrino masses, but under the constraint that its total value is $\sum_i m_{\nu_i} = 4.6$ eV, which corresponds to $\Omega_\nu = 0.2$ (for $h = 0.5$). In the CHDM models with slightly tilted primordial spectra (Pogosyan & Starobinsky 1995a) or with more than one massive neutrino (Pogosyan & Starobinsky 1995b), the range of $(1 - \Omega_{CDM}, h)$ parameters is slightly broader than for the standard CHDM model with one neutrino. However, the allowed value for the Hubble constant is $h \leq 0.6$, which still might be too small. Note that for the $\Lambda$CDM model the allowed parameters are given by $\Omega h \approx 0.25$ and $\Omega > 0.3$ (Kofman, Gnedin & Bahcall 1993; Ostriker & Steinhardt 1995).

To study the viability of the CHDM model against the accumulating observations of clusters, such as X-ray emission, strong lensing producing arcs and weak lensing producing background galaxies distortion, and galaxy velocity dispersion, one needs to know the mass and velocity dis-tributions in the dark matter components in the clusters. In this paper we study the density and velocity profiles of clusters in the CHDM cosmological models. We will make calculations primarily for the CHDM models with one 4.6 eV or two 2.3 eV neutrinos. However, we would like to note that the physical effects we are discussing here are more general and are applicable to a broad range of the CHDM-type models. We also compare the results with the corresponding predictions in the $\Lambda$CDM model, and make preliminary comparisons with some observations of clusters.

It is well known that the properties of clusters such as the mass density profile $\rho(r)$ and the (galaxy) velocity dispersion profile $\sigma_v(r)$ can probe the background cosmological model, as well as the index $n$ of the power spectrum $P(k)$ (see, for example, Gunn 1972; Hoffman & Shaham 1985; Quinn, Salmon, & Zurek 1986; Crone, Evrard, & Richstone 1994 and references therein). For hierarchical models like CDM, the formation of a cluster can be treated, at least qualitatively, as a spherical accretion onto a peak of the initial density field. For the peak one can take, say, a massive galaxy with mass $10^{13} M_\odot$ (Gunn 1972). The accretion model gives us insight into how the density and velocity profiles in the cluster outskirts depend on the parameters of the cosmological model like the CDM model or its interesting variations.

However, in the CHDM model the process of clusters formation is more complicated due to the nature of the neutrino component. Indeed, hot neutrinos accrete onto the protocluster only after their thermal velocities $v_\nu$ drop below the velocity dispersion of the protocluster. Thus, the neutrino halo of the cluster starts to form after the cold dark matter halo is well advanced into the nonlinear stage. Neutrinos also cannot be compressed to an arbitrarily high density in the cluster center. Because neutrinos are collisionless particles, their phase-space density is conserved. This puts a limit on their mass density in gravitationally bound objects. The sizes of the objects that should have the neutrino halo are related to the neutrino masses. For the pure hot dark matter model the phase space density constraint leads to the Tremaine & Gunn (1979) criterion, which links the mass of the neutrino to the parameters of the isothermal halo: the 24 eV electron neutrinos compose the halo of objects of the size of normal galaxies. One expects that in the CHDM model, a few eV neutrinos do not contribute to masses of galaxy halos, but they can make a contribution to masses of clusters. In this paper we extend the Tremaine-Gunn criterion for the CHDM models and will apply it for the clusters of galaxies.

In this paper we also address two specific questions of the cluster structure for the CHDM-class models. How small is the fraction of the neutrino mass density, $f = \rho_\nu/\rho_{CDM}$, in the core of a cluster? This question is rather topical, because it is related to the problem of the baryon-to-dark-matter ratio in clusters. We calculate the compression factor $f$ as function of the neutrino masses. Next, what is the profile of the neutrino mass density in the halo of clusters relative to that of CDM? This question is related to other important problems: the potential discrepancy between the total surface mass density which can be reconstructed from weak gravitational lensing, and the X-ray emission from clusters.

In Sec. 2 we study the differences in the evolution in the linear regime between the cold dark matter and the neutrinos. In Sec. 3 we derive the phase space density constraint on the neutrino density for virialized clusters. In Sec. 4 we derive the cold particles and neutrino density profiles in clusters, first analytically and then by N-body simulations. In Sec. 5 we discuss some observational constraints on the baryonic, cold dark matter and neutrino components in clusters. First we derive the temperature profile of the hot gas in the hydrostatic equilibrium with the CHDM mass density and compare it with the data. Then we obtain the total surface density in the CHDM model which is probed by the weak gravitational lensing. Finally we discuss the galaxy velocity dispersion profile in clusters as a discriminative test for cosmological models.

## 2 Linear Fluctuations in CDM and Neutrinos Components

In a model with only a cold dark matter component, the formation of a protocluster can be roughly viewed as an infall of spherically symmetric shells onto a local maximum of the primordial field of gaussian density perturbations $\delta(r)$. In order to study some statistical aspects of cluster formation, such as cluster abundance with redshift, the primordial density field is smoothed with a filter, with a scale corresponding to the cluster mass, $\sim 10^{15} M_\odot$. Then the cluster abundance can be estimated via the statistics of the overdense regions of the filtered $\delta$-field above some threshold $\delta_c$, as in the Press-Schechter model. In this paper we address instead the shape of density profile of clusters. For this purpose we consider the initial density peaks relevant for the protoclusters, and smooth the initial density field on that scale, which corresponds to a massive galaxy with mass $10^{13} M_\odot$. The structure of a cluster halo forming by the accretion onto the initial overdensity is then defined by the details of the density profile around the maximum. The later can be well described by the two-point linear correlation function $\xi(r) = <\delta(r)\delta(0)>$ (e.g., Hoffman 1988).

Specific features of cluster formation in the CHDM model with cold and hot components can be calculated in the linear regime. Let us consider the density profiles in both components around the density peak. Linear evolution of the adiabatic perturbations in the CHDM model is derived from the kinetic equation for neutrinos and the linearized Einstein equations (see Appendix A). Solving these equations numerically, one can calculate the transfer function of the total density fluctuations, which is $C(k) = k^{-1} P(k)$ for the scale-free initial power spectrum. Figure 1 shows transfer functions $C(k)$ at $z = 0$ for three CHDM models: only one neutrino is massive, $m_\nu = 4.6$ eV (upper curve); two neutrinos have equal masses 2.3 eV (middle curve), and all three neutrinos have equal masses 1.53 eV (lower curve). The upper and the lower curves are the limiting cases. Curves corresponding to all other distributions of 4.6 eV among three types of neutrinos will be sandwiched between these limits. This rule is valid not only for the transfer function but for other characteristics we will consider throughout the paper (see also Pogosyan & Starobinsky 1995b). The splitting of $C(k)$ for different masses on clusters scales is quite noticeable.

The overdensity profiles in cold ($A = 1$) and hot ($A = 2$) components around the peak can be expressed through the corresponding correlation functions, $\delta_A(r) = \nu \xi_A(r)/\sigma_A$, where $\nu$ is the peak height in units of the $rms$ density dispersion $\sigma_A$. The density profile is simply $\rho_A(r) = \Omega_A \bar{\rho}(1 + \delta_A(r))$. From the linear equations of Appendix A we numerically calculate the two-point correlation functions for both components, as functions of the filtering mass $M$ and redshift $z$. For definiteness, we choose the model with two 2.3 eV neutrinos. In Figure 2 we plot average CDM (solid curve) and 2.3 eV neutrino (dashed curved) density profiles $\rho_A(r)$ around a density peak; in upper panels for filtering on the scale $10^{13} M_\odot$ (with the peak heigh $\nu = 1.68$), in lower panels after filtering on the scale $10^{15} M_\odot$; in the left panels at $z = 4$, in the right panels at at $z = 0$. The difference in the evolution of the overdensities at a given scale depends on how that scale relates to the neutrino Jeans length $\lambda_J$:

$$\lambda_J \approx 3h^{-1} \left(\frac{2.3eV}{m_\nu}\right)(1+z)^{-1/2} Mpc. \tag{1}$$

This corresponds to the Jeans mass $M_J \sim 10^{13} M_\odot$ at $z = 4$ (for 2.3 eV neutrinos). For the upper left panel in Figure 2 the scale of the fluctuations is less than $\lambda_J$. As the result, the growth of the CDM fluctuations at that scale is significantly advanced relative to the neutrino fluctuations. Compare this with the evolution shown in the lower panels at scales greater than $\lambda_J$, where the growth in both components are comparable. Therefore the formation of clusters begins with the formation of the CDM protoclusters.

At early stages hot neutrinos do not accrete on the CDM protoclusters because the typical neutrino velocity $v_\nu$ is too large:

$$v_\nu \approx 70(1+z)\left(\frac{2.3\text{eV}}{m_\nu}\right) km/s. \tag{2}$$

Note that $\lambda_J \sim \int dt \cdot v_\nu$. Neutrinos accrete on the protocluster only after $v_\nu$ drops below the velocity dispersion of the protocluster. At present, neutrinos have accreted into the halos of clusters. This problem is reminiscent to the neutrino accretion on the nonperturbative density inhomogeneities (seeds) formed by cosmic string loops (Brandenberger, Turok & Kaiser 1986). If one models a seed as a delta-function overdensity, then neutrinos form an extended halo on the scale $\sim \lambda_J$ around it.

One might think that in the CHDM model, there would be an extended neutrino halo, where neutrinos dominate over cold particles, around a cluster where cold particles dominate over neutrinos. This is not true. In the CHDM model neutrinos make a contribution to the cluster halo, but their density nowhere dominates over the CDM density. As we will see from N-body simulations in Sec. 4, the ratio of the neutrinos density to the CDM density never exceeds the background ratio 1/4. This result can be understood if we consider the spherical infall model implemented for the neutrinos whose thermal velocity is already much lower than the clusters velocity dispersion.

## 3  Neutrino Phase Space Density Constraint

For clusters in equilibrium, a useful constraint on the density of the neutrinos inside massive halos can be derived from the distribution of neutrinos in phase space. In this and the next Sections we extend the Tremaine & Gunn (1979) criterion to the CHDM model. The constraint comes from the condition that for the collisionless particles the phase space density $n(\vec{p}, \vec{x})$ is constant along the particles trajectories. The initial phase space density of neutrinos of one family $\nu_i$ is given by:

$$n_{init}(\vec{p}, \vec{x}) = \left(e^{pc/kT_\nu} + 1\right)^{-1}, \tag{3}$$

where $T_\nu$ is the temperature of neutrinos and $p$ is their momentum. The initial phase space density distribution $n_{init}$ has the maximum value $n_{init,\max} = 0.5$ at $p = 0$. Thus by equating the final phase space density with the maximum of the initial phase space density we get upper limit on the density in the final distribution.

Let $n_f(\vec{p}, \vec{x})$ be the final phase space density of neutrinos of one family $\nu_i$ gravitationally bound in the massive spherically symmetric halo. The space density of neutrino $\nu_i$ at radius $r$ is the integral of $n(\vec{p}, \vec{x})$ over the momentum $\vec{p} = m_{\nu_i} \vec{v}$

$$\rho_{\nu_i}(r) = \frac{2m_{\nu_i}}{(2\pi\hbar)^3} \int_0^\infty d^3p \; n_f(\vec{p}, r). \qquad (4)$$

Let the one-dimensional velocity dispersion in the halo be $\sigma$, then in (4) we can factor out all dimensional parameters

$$\rho_{\nu_i}(r) = \frac{m_{\nu_i}^4 \sigma^3}{\pi^2 \hbar^3} \int dy\, y^2 n_f(y, r), \qquad (5)$$

where $y = v/\sigma$. Now we can estimate the maximum value of the neutrino $\nu_i$ mass density at the center $\rho_{\nu_i}(0)$. The final phase space density is assumed to depend only on the energy

$$n_f(\vec{p}, \vec{x}) = n_0 \cdot f(E), \qquad (6)$$

where $n_0$ is a normalization constant, and $E$ is the particle energy: $E = y^2/2 + \phi(r)/\sigma^2$. The gravitational potential $\phi(r)$ is normalized at the minimum to $\phi(0) = 0$, and the distribution function $f(E)$ is normalized to $f(0) = 1$. The integral in (5) for common distributions $f(E)$ (including those we will use below) can be evaluated as $\int dy\, y^2 f(y^2/2) \simeq O(1)$; for example, for the Boltzmann distribution $\int dy\, y^2 f(y^2/2) = 1.25$. Below we will distinguish the fine- and coarse- grained distributions, see e.g. Binney & Tremaine (1987). The coarse-grained distribution $< n(\vec{p}, \vec{x}) >$ is the fine-grained distribution averaged over some phase volume; when the pair relaxation is negligible that is always lower than the actual, fine-grained space density.

The phase space density is invariant along the neutrino trajectories. Therefore the final fine-grained distribution (6) cannot exceed the maximum value $1/2$ of the initial distribution (3), $n_0 < 0.5$. Thus, an upper limit on the neutrino $\nu_i$ density in the halo center is

$$\rho_{\nu_i}(0) \leq \frac{m_{\nu_i}^4 \sigma^3}{2\pi^2 \hbar^3}. \qquad (7)$$

We can give not only the upper limit but also a reasonable estimation of the central neutrino mass density. An average (coarse-grained) initial phase space density $< n(\vec{p}, \vec{x})_{init} >$ can be found either from the mean hot neutrino energy, or from the median hot neutrino energy, or from the condition that half of all neutrinos exceeds the value $< n(\vec{p}, \vec{x})_{init} >$. All these methods give us about the same value $< n(\vec{p}, \vec{x})_{init} > \approx 0.05$. By equating the final average phase space density to this value, we obtain the expected neutrino density at the center

$$\rho_{\nu_i}(0) \simeq 0.1 \cdot \frac{m_{\nu_i}^4 \sigma^3}{2\pi^2 \hbar^3}. \qquad (8)$$

Using the equation (8), we can evaluate the neutrino mass density at the center of clusters for different neutrino masses. In the model with a single massive neutrino with mass 4.6 eV, we have $\rho_\nu(0) \simeq 2 \cdot 10^{-26} \, (m_\nu/4.6\text{eV})^4 \, (\sigma/1000 km/s)^3 \, g/cm^3$; in the model with two massive neutrino with mass 2.3 eV each, the total central neutrino mass density is $\rho_\nu(0) = 2\rho_{\nu_i}(0) \simeq 2.5 \cdot 10^{-27} \, (m_\nu/2.3\text{eV})^4 \, (\sigma/1000 km/s)^3 \, g/cm^3$; in case of three massive neutrinos with mass 1.53 eV each, total central neutrino density is $\rho_\nu(0) = 3\rho_{\nu_i}(0) \simeq 7.4 \cdot 10^{-28} \, (m_\nu/1.53\text{eV})^4 \, (\sigma/1000 km/s)^3 \, g/cm^3$. The neutrino mass density at the center strongly depends on the neutrino mass. However, it is only by factor $10^2 - 10^3$ greater than the mean density $\bar\rho = 1.88 \cdot 10^{-29} h^2$ g/cm$^3$. The CDM component, filtered with the scale of a big galaxy, has at the center $\rho_{CDM} \simeq 3 \cdot 10^4 \bar\rho$. Thus, at the center, the CDM component significantly dominates over the HDM one.

## 4 CDM and Neutrinos Halo Density Profiles

We study the neutrino density profile $\rho_\nu(r)$ together with the CDM density profile $\rho_{CDM}(r)$ in the clusters. An especially interesting characteristic is their ratio $f(r) = \rho_\nu(r)/\rho_{CDM}(r)$, which describes the relative compression of the cold particles in the clusters. First, we use an analytic model based on certain assumptions about the neutrinos and cold particles phase space distributions. Then we compare our predictions with results of N-body simulations for the CHDM model.

### 4.1 Analytical Model

Finding the mass density profiles of the two components requires a generalization of the one (neutrino) component problem considered by Tremaine & Gunn 1979. To derive the neutrino density profile $\rho_\nu(r)$ together with the CDM density profile $\rho_{CDM}(r)$ in the clusters, in the first turn, one should specify the phase space density distributions in both components.

A convenient model for the CDM particles is the spherically symmetric isothermal distribution

$$n_{CDM}(\vec{v}, \vec{r}) = N \cdot \exp\left(-\frac{E}{\sigma_{\text{cold}}^2}\right), \tag{9}$$

where $E = v^2/2 + \phi(r)$, $\sigma_{\text{cold}}$ is the velocity dispersion of cold particles and $N$ is a normalization constant. For this phase space density distribution, the mass density distribution of the CDM particles is

$$\rho_{CDM}(r) = \rho_{CDM}(0) \cdot \exp\left(-\frac{\phi(r)}{\sigma_{\text{cold}}^2}\right), \tag{10}$$

where $\rho_{CDM}(0)$ is the central CDM density.

We take the following distribution function $f(E)$ for neutrino (see eq. (6)):

$$n(\vec{p}, \vec{x}) = n_0 \cdot \frac{1 + \exp(-\mu)}{1 + \exp(E/\sigma_\nu^2 - \mu)}, \tag{11}$$

where $\sigma_\nu^2$ is the neutrino velocity dispersion, and $\mu$ plays the role of the chemical potential, which we treat as a free parameter and will be fixed below. Note that the cold particles velocity dispersion $\sigma_{\text{cold}}$, generally speaking, is different from $\sigma_\nu$, but here we assume they are equal. Using (11) in the formula (5), we obtain the neutrinos mass density $\rho_{\nu_i}(r)$, see Appendix B for technical details. At this point the parameter $\mu$ can be fixed by considering the ratio $\rho_\nu(r)/\rho_{CDM}(r)$ at large distances $r$ where it should match the background ratio $\Omega_{\nu_i} : \Omega_{CDM}$. Then the neutrino density profile $\rho_{\nu_i}(r)$ is the function of the gravitational potential $\phi(r)$ and a parameter $\rho_{CDM}(0)$ only, see eq. (27) of Appendix B.

Now we are ready to write down the Poisson's equation for the gravitational potential $\phi(r)$ for the two component spherically symmetric configuration

$$\frac{d^2\phi}{dr^2} + \frac{2}{r}\frac{d\phi}{dr} = 4\pi G \left(\rho_{CDM}(r) + \rho_\nu(r)\right), \qquad (12)$$

where the CDM and neutrino densities as functions of $\phi(r)$ are given by eqs. (10) and (27). In Appendix B we reduce the Poisson equation to the form of a generalized Emden equation for a selfgravitating isothermal distribution. We solve numerically the Emden equation for the two component halo for given parameter $\rho_{CDM}(0)$. We have to choose the value of $\rho_{CDM}(0)$. Generally speaking, how large can this value be is still an open question, since for the CDM models density has $1/r$ cusp up to the scale of resolution (Dubinsky & Carlber 1991). Our numerical simulations have a resolution of a large elliptical galaxy. Therefore as a central CDM density in clusters we will mean the "coarse-grained" density. We choose $\rho_{CDM}(0)$ equal to $3 \cdot 10^4 \bar{\rho}$, the average density of a large elliptical galaxy.

In Figure 3 we plot the final results in terms of the ratio of the two density profiles $f(r) = \rho_\nu(r)/\rho_{CDM}(r)$, for three cases: one 4.6 eV, two 2.3 eV and three 1.54 eV neutrinos. We see that in the case of two 4.6 eV neutrinos the neutrino mass density in the core is suppressed by a factor of about 2 relative to the CDM density. In the case of one 2.3 eV neutrinos the suppression factor is about 5. In the case of three 1.54 eV neutrinos the suppression factor is about 10. At distance about 1 Mpc, neutrinos contribute to the cluster outskirts in the proportion $\Omega_\nu : \Omega_{CDM}$. In the limit of large $r$, the isothermal analytical model, obviously, gives for both densities the asymptote $\propto r^{-2}$, which is, as we will see, is slightly flatter than the actual dependence $\propto r^{-2.5}$, which we find in numerical simulations. However, the analytical model predicts the ratio of the two density profiles fairly accurately comparing with that from numerical simulations (upper panel of Fig. 4).

## 4.2  Numerical Simulations

To obtain detailed density structure of clusters in the CHDM model, we made an N-body simulation. In addition to the density profiles, we use the simulation to get the profiles of the velocities in the clusters. The simulation has been made in a box of $50h^{-1}$ Mpc size with periodic boundary conditions. It has $256^3$ cold particles and twice that many hot particles moving in a $800^3$ mesh. The (formal) resolution is $62.5h^{-1}$ kpc. Initial conditions correspond to a COBE normalized ($Q_{\text{RMS-PS}} = 18\mu$K) CHDM power spectrum with total $\Omega = 1$, $\Omega_{CDM} = 0.8$, $\Omega_\nu = 0.2$, and $h = 0.5$. We chose the CHDM model with two equal mass 2.3 eV neutrinos. We also will compare the results with those for the "old" 7 eV model. For this case we use another simulation with the same number of particles and box size, but with slightly worse resolution ($512^3$ mesh).

Two largest clusters – (a) and (b) – in the simulation for the model with two neutrinos were chosen as examples. They are smaller, but comparable to the Coma cluster. The cluster (a) has mass $6.5h^{-1}10^{14}M_\odot$ (left panel in Figure 4) and cluster (b) has mass $5.5h^{-1}10^{14}M_\odot$ (right panel in Figure 4) within the Abell radius $1.5h^{-1}$Mpc. In the lower panel of Figure 4 we plot the corresponding CDM density profiles. Now we discuss the fit by simple analytic form for the CDM density profile. There are uncertainties in the fit due to the numerical resolution of our simulations: the very central part of the clusters is not resolved.

Navarro, Frenk, & White (1995) found that for the CDM model the dark-matter density profile in central parts of clusters is well approximated by $\rho(r) = 60\bar{\rho}r_v^3/r(r + 0.2r_v)^2$, where $r_v$ is the virialized region the cluster (radius at which the mean overdensity is equal to 200). We find that this formula also gives a very good fit to the CDM density profile in our CHDM models inside central $\approx 1h^{-1}$ Mpc region. But at larger radii the density profile in CHDM models is shallower: it declines as $\rho \propto r^{-\alpha}$, where $\alpha$ is usually less than 3. To fit the density profile both within and outside cluster, we will use a simple conjecture which simultaneously gives us $1/r$ asymptote at the center, and $1/r^\alpha$ asymptote on the cluster outskirt

$$\rho_{CDM}(r) = \frac{C}{r(r+R)^{\alpha-1}}. \qquad (13)$$

This density profile gives a better fit than the King-like profile, although the later is also quite reasonable. The profile (13) provides a central surface mass profile large enough to produce arcs from the strong gravitational lensing. Equation (13) has been tested for several clusters in the simulation. We find the slope in the halo for the model with two 2.3 eV neutrinos to be $\alpha \approx 2.5$. The same slope $\approx 2.5$ in the halo was earlier found for clusters in the CDM+$\Lambda$ model (Crone, Evrard & Richstone 1994). Thus, clusters CDM density profiles slopes in the 2.3 eV CHDM and CDM +$\Lambda$ models are similar. The value of the scale parameter $R$ in (13) we found is $R \simeq 200 - 300$ kpc (for $h = 0.5$). The neutrino density profile is fitted by the formula $\rho_{nu}(r) = f(r)\rho_{CDM}(r)$, where $f(r)$ is the compression factor, plotted in Fig. 3.

In Figure 5 we plot the CDM density profiles in the "old" CHDM model with $m_\nu = 7$ eV. In this model the density profile slope is steeper, $\alpha \approx 2.15$. It is interesting that the slope $\alpha$ is rather sensitive to the details of the CHDM models, such as $\Omega_\nu$ and the spectrum of the neutrino masses. Partly the difference in $\alpha$ is caused by different power spectrum index $n$ at the clusters scale. For $\Omega_\nu = 0.2$, 2.3 eV neutrinos, $n \approx -1.36$; for $\Omega_\nu = 0.3$, 7 eV neutrinos, $n \approx -1.56$. The upper panels of Figure 4, 5 show the ratio $f(r)$ of hot-to-cold densities as a function of distance from the cluster center, normalized in such a way that unity corresponds to the cosmological ratio $\Omega_\nu : \Omega_{CDM}$. We see that indeed the neutrino density at the center is suppressed by a factor 5 relatively to the CDM density in the case of 2.3 eV neutrinos (Figure 4); and by factor 2 in the case of 7 eV neutrinos, in agreement with the analytic prediction, cf. Figure 3. At about $\sim 1h^{-1}$ Mpc the ratio $f(r)$ saturates and reaches its background value $\Omega_\nu : \Omega_{CDM}$. The fluctuations of the ratio $f(r)$ at large $r$ are not statistical. Both clusters (a) and (b) have neighbors – smaller clusters at about $1h^{-1}$ Mpc from the cluster (a) and $2h^{-1}$ Mpc from cluster (b). As the result, the minima of $f(r)$ correspond to the positions of other clusters and groups.

What appears to be most interesting is the profile of the velocity dispersions $v_{rms}(r)$ in the clusters. Figure 6 shows $v_{rms}(r)$ for clusters (a) and (b) for the 2.3 eV model. The solid curve shows $v_{rms}(r)$ of cold particles; the dashed curves are for hot particles. Outside of the Abell radius both curves are close one to another. At smaller radii, neutrino velocities are larger than CDM velocities. Note that $v_{rms}(r)$ of cold particles, which (to some extent) mimics the profile of the galaxies velocity dispersion in the clusters halo, stays almost constant, at least up to distance $1.5h^{-1}$ Mpc. We define a *local* velocity dispersion $\sigma_{rms}(r)$, which mimics the local "temperature". Because there could be bulk flows inside the clusters halo, some fraction of the peculiar velocities will not contribute to temperature. Let us define the bulk flow velocity $v_f(r)$ as the mean velocity of cold particles within a sphere of radius $150h^{-1}$ kpc centered on the each cold particle. Then the local velocity dispersion is $\sigma_{rms}(r) = <(v - v_f(r))^2>^{1/2}$. The profile of $\sigma_{rms}(r)$ is shown as the dash-dotted curves in Figure 6. The difference between the two upper curves and lower curve $\sigma_{rms}(r)$ is a measure of the departure from hydrostatic equilibrium in the cold and hot components. Figure 6 indicates that the hydrostatic equilibrium is a reasonable assumption within about radius $\sim 1h^{-1}$ Mpc, but fails significantly at larger radii. It can be well explained by radial infall into the cluster halo. Also note that the cluster (a) is more distorted by the bulk flows, associated with its close neighbor as compared to the cluster (b).

# 5 Observational Tests on Clusters Density and Velocity Profiles

In the CHDM model there are three matter components, cold, hot and baryonic, which are unequally distributed within the cluster. The baryonic component can be probed with the X-ray emission of intracluster gas. The distribution and motion of galaxies are primarily signatures of the cold component. The weak gravitational lensing would be produced by the total gravitational potential of the cluster. Therefore a combination of all three sorts of data would provide a test for the CHDM model. In this section we discuss some expected properties of clusters in the CHDM model regarding these observational tests.

## 5.1 X-ray Surface Brightness and Temperature Profiles

In this Section we use a simple model to check to which extend the dark matter density fit (13) is compatible with the X-ray data for clusters. By all means, this does not replace the numerical simulations with dark matter and gas, however, this is an instructive estimation. We will use, for simplicity, the model hydrostatic equilibrium of hot gas in the spherically symmetric static gravitational potential produced by the dark matter with the density profile (13). Then we will use well-known X-ray surface brightness profile for the cluster A2256, to calculate a theoretical temperature profile. An actual temperature profile is compatible with a constant, the departure of the theoretical temperature from the constant measures the accuracy of this modelling.

Let us assume that the gas density is spherically symmetric and is given by the $\beta$-model (Cavaliere & Fusco-Femiano 1976)

$$\rho_{gas}(r) = \rho_{gas}(0) \left[1 + \left(\frac{r}{a}\right)^2\right]^{-3\beta/2}, \tag{14}$$

where $a$ is a core radius. Then the X-ray surface brightness, $\Sigma_X(r)$, is

$$\Sigma_X(\theta) = \Sigma_X(0) \left[1 + \left(\frac{\theta}{\theta_a}\right)^2\right]^{-3\beta+1/2}, \tag{15}$$

where $\theta_a$ is the angular size corresponding to $a$. The X-ray surface brightness of clusters typically is far from being spherically symmetric. However, the azimuthally averaged surface brightness can be fitted very well by formula (15). In Figure 7a we plot, as an example, $\Sigma_X(\theta)$ for the Abell cluster A2256 (Briel & Henry 1994). This object is a rich, symmetric, partially relaxed cluster, similar to the Coma cluster. The best fit for the parameters $\beta$ and $a$ are $\beta = 0.81 \pm 0.01$, and $a = 0.27h^{-1}$ Mpc. Note a very high accuracy $\sim 1\%$ in determining $\beta$.

Assuming gas is in steady state hydrostatic equilibrium with the spherically symmetric binding cluster gravitational potential, we have (Sarazin 1986)

$$\frac{4\pi}{r}\int_0^r dr' r'^2 \rho(r') = -\frac{kT(r)}{\mu m_p G}\left(\frac{d\ln\rho_{gas}(r)}{d\ln r} + \frac{d\ln T(r)}{d\ln r}\right), \tag{16}$$

where $\mu m_p$ is the mean particle mass in the gas, and $T(r)$ is the gas temperature. This formula, in principle, relates the total dark matter mass density $\rho(r)$, given by (13), to the gas density (14). Unfortunately, even this oversimplified model based on (16) requires the knowledge of the temperature profile $T(r)$, in order to reconstruct the dark matter density $\rho(r)$. The measured X-ray temperature profile $T(r)$ at present is known with a low accuracy, see Figure 7b. The actual temperature is compatible with the constant $\approx 7$ Kev. The usual strategy is to insert the dark matter density in a chosen form (for instance, in the form (13)) into (16), and solve it for $T(r)$ as function of the parameters of dark matter density profile. (Hughes 1989, Briel, Henry & Böhringer 1992). One finds from (16)

$$T(r) = \rho_{gas}(r)^{-1}\left(T(0)\rho_{gas}(0) - \frac{4\pi\mu m_p G}{k}\int_0^r \frac{dr'}{r'^2}\rho_{gas}(r')\int_0^{r'} dr'' r''^2 \rho(r'')\right). \tag{17}$$

The resulting $T(r)$ profile is compared with the observed data to constrain parameters of the density profile. Despite the low accuracy of the measured temperature, this test turns out to be sensitive to the modelling of the dark matter profile. Our approach is complementary to the method of reconstruction of the X-ray surface brightness by adjusting the dark matter profile under the assumption that the actual temperature is constant, which is compatible with observations (Miralda-Escude & Babul 1995).

Substituting the expressions (13) and (14) in (17), and using their asymptotes at large $r$ it is easy to see that the integral in the right hand side converges to some constant $J$, which depends on $R, \alpha, a, \beta$ and the dark matter density at the center $\rho_0$. The analysis shows that the integral is rapidly saturated at $r \sim 0.5$ Mpc, since the expression in the bracket goes rapidly to the constant $I = T(0)\rho_{gas}(0) - J$. Therefore at larger $r$ the temperature $T(r)$ crosses zero if $I < 0$, or increases very fast as $\rho_{gas}^{-1}(r)$ if $I > 0$. A physically acceptable temperature profile can be achieved only if $T(0)\rho_{gas}(0)$ is tuned to be close to $J$ with an accuracy better than 1%.

The projected temperature is

$$T(R) = 2\sigma_{gas}(R)^{-1} \int_R^\infty \frac{dr\, r}{\sqrt{r^2 - R^2}} T(r)\rho_{gas}(r)^2, \qquad (18)$$

where $\sigma_{gas}(R)$ is the projection of the value $\rho_{gas}^2(r)$. The theoretical temperature profile reduced from this formula for the chosen parameters of densities in gas (14) and total dark matter (13) components is plotted at Figure 7b. The data points correspond to the actual temperature measurement for the A2256 (Briel & Henry 1994). The theoretical prediction for $T(r)$ from the CHDM model are compatible with the data for the A2256 cluster outside the radius $\sim 0.5$ Mpc, but departures from the data up to factor 2 at the center. The temperature spike around the center depends on the details of the dark matter density peak. Traditional King profile for the dark matter density with the parameters adjusted to the numerical results of Fig. 4 gives much bigger departure of $T(r)$ from the constant. Thus, either the model of hydrostatic equilibrium or the model (13) of the spherically symmetric dark matter distribution should be modified.

## 5.2 Mass Density Surface Profile from Lensing

The method, based on the weak gravitational lensing, can in principle, reconstruct the total projected surface density $\Sigma(r)_{tot}$. In the CHDM model, halos of galaxies are made from the cold dark matter particles. We expect therefore that in clusters, galaxies are tracers of the cold dark matter. From the known cold dark matter density profile, one can obtain the projected surface density of cold dark matter

$$\Sigma(R)_{CDM} = 2 \int_R^\infty \frac{dr\, r}{\sqrt{r^2 - R^2}} \rho_{CDM}(r). \qquad (19)$$

In the same way one can obtain the total projected surface density $\Sigma(r)_{tot} = \Sigma(r)_{CDM} + \Sigma(r)_\nu$.

To see the difference in $\Sigma(r)_{tot}$ and $\Sigma(r)_{CDM}$, In Figure 8 we plot $\Sigma(r)_{CDM}$ as dashed curve, and $\Sigma(r)_{tot}$ as the solid curve for the CHDM model with 2.3 eV neutrinos, calculated from formulae (19), (13) and $f(r)$ from simple analytical model of Sec. 4.1. The lesson is that both total and CDM surface densities have almost similar shapes, the amplitude of the total surface density exceeds the CDM one by about 10%. The presence of the neutrino halo in the clusters basically changes only the amplitude of the surface density. Both uncertainties in the galaxies surface densities (Kaiser & Square 1993) as well as in the reconstructed surface density from the lensing (Fahlman et al 1994) make the neutrino halo of clusters in the flat CHDM model with $\Omega_\nu = 0.2$ virtually undetectable by this method.

However, in the framework of the CHDM model one can calculate the details of the mass density surface profile in the model. In Fig. 8 note the asymptotic slope $\Sigma(R)_{tot} \propto R^{-1.5}$ at large distance from the center, the asymptote $\Sigma(R)_{tot} \propto log R$ at small $R$, and the bend of the surface profile at $R \sim 0.1$ Mpc. Also note the remarkable difference between the hot gas mass density in Figure 7a and that of dark matter in Figure 8. In the $\Lambda$CDM model the density profile is steeper inside the cluster, therefore one expect the bend of the mass surface profile is at smaller $R$. The asymptotic slope $-1.5$ is the same. The clusters in the $\Lambda$CDM model are more relaxed than those in the CHDM model. Thus, the details of the surface mass density $\Sigma(\vec{R})_{tot}$ might be another test to distinguish cosmological models.

## 5.3 Cluster Galaxy Velocity Dispersion Profile

The velocity dispersion $\sigma_v(r)$ of galaxies as function of the radius has a significantly different profile in different models. As we described in Sec. 4, numerical simulations in the CHDM model show that $\sigma_v(r)$ is almost constant in the tested range of scales up to distance $\sim 3h^{-1}$ Mpc. The projected velocity dispersion is constructed from $\sigma_v(r)$ by the formula similar to (18), and therefore also close to the constant. On the contrary, the $\Lambda$CDM model as well as the open CDM model show $\sigma_v(r)$ decreasing with radius (c.f. Crone et al 1994).

The observed projected galaxy velocity dispersion, apparently, has a central maximum, and then decreases with distance from the cluster center (e.g. Kent & Gunn 1982; Sharpes, Ellis & Gray 1988). Potentially, this is very interesting test because this is in the conflict with the CHDM prediction, and but not with $\Lambda$CDM model. However, for more definite conclusion we have to take into account the complicated observational procedure of the rejecting of interlopes. The CNOC cluster redshift survey indicates a smaller fall of the projected velocity dispersion up to $\sim 1.5h^{-1}$ Mpc (Yee et al 1994). (Note that this survey covers the range of clusters redshift $0.2 \leq z \leq 0.5$, where cosmological constant in the $\Lambda$CDM model has no significant cosmological impact). Thus, more data on the velocity dispersion profiles in clusters on different redshifts provide a discriminating test of cosmological model.

# 6 Discussion

We have studied several effects, responsible for cluster formation and structure in the Mixed Dark Matter models. Neutrinos in the range of masses up to a few eV contribute to the halo of clusters of galaxies, but not to that of galaxies or group of galaxies. Protoclusters form first from the cold dark matter particles, hot neutrinos are accreted afterwards onto the CDM protoclusters. Cold dark matter dominates most of the cluster mass density. The neutrino mass density is especially suppressed relative to the CDM density at the cluster center. At a distance of $1h^{-1}$ Mpc the ratio of densities of both components saturates at the background ratio. Thus, only $\simeq 80\%$ of dark matter is clustered inside the Abell radius. Therefore in the CHDM model the ratio of baryons to the dark matter inside the clusters should be in 1.25 times greater than the background ratio. This is not enough to resolve the problem of large baryon fraction in clusters (Briel, Henry & Böhringer 1992; White et al 1993). In the model with a cosmological constant, the ratio of baryons to the cold dark matter inside the clusters is about 3 times greater than the cosmological ratio of baryons to the total dark matter, including $\Lambda$-term. Thus, there is no baryonic catastrophe problem in the $\Lambda$CDM model.

We found the density and velocity profiles of cold particles and neutrinos in clusters. The slopes $\alpha \simeq 2.5$ of the halo density profiles in both components are similar. The exact value of $\alpha$ depends on the neutrino mass. It should be noted that from the density profile slope alone one cannot distinguish the CHDM model, the $\Lambda$CDM model, or tilted CDM models. On the other hand, the velocity profile in the CHDM model is almost constant, while it is falling off with radius in the $\Lambda$CDM model. The projected velocity dispersion in clusters is a strong test to distinguish cosmological models.

The oversimplified model of the spherically symmetric hot gas distribution in the hydrostatic equilibrium in the CHDM clusters was used to calculate the X-ray theoretical temperature $T(r)$ from given X-ray surface brightness $\Sigma_X(\theta)$ and spherically symmetric dark matter density profile. Around the center the theoretical temperature profile departures from the constant compatible with the data on A2256 cluster.

The total mass density in the clusters outskirts has a $\sim 20\%$ excess relatively to that inside the cluster, due to the neutrino contribution. Galaxies in clusters, presumably, are good tracers of the cold dark matter component. The weak gravitational lensing would be a probe of the total cluster mass, due to cold dark matter and neutrinos. We found that the total projected mass density differs from the cold dark matter projected density by amplitude but not by the shape. The weak lensing better reconstructs the shape of the projected density. The slope of the total projected mass density of the clusters predicted by the model is $\simeq -1.5$. The same slope is predicted in the $\Lambda$CDM model. The weak lensing method in principle, could probe the model by more subtle statistical tests of the cluster surface mass distribution.

In summary, our study of the cluster properties in the CHDM model are adding several tests of the model.

We thank S. Tremaine for useful discussions of the neutrino phase space density form. L.K. thanks the support from IFA, University of Hawaii, D.P was supported by CITA and CIAR cosmology program, and partly by the ITP UCSB. A.K. thanks IfA, University of Hawaii, for hospitality and the NSF grant AST-9319970. P.H. was supported by the NSF grant AST 9119216 and AST 950515 and NASA grant NAG5-1880.

# APPENDIX A: Linear Fluctuations in the CHDM model

We have calculated the linear evolution of adiabatic perturbations in the model, treating neutrinos kinetically through the Vlasov equation, and the CDM component as dust. We describe small scalar perturbations around a homogeneous, isotropic flat background universe in the form

$$ds^2 = a(\eta)^2 \left[ (1 + 2\Phi)d\eta^2 - (1 - 2\Psi)dx_\alpha dx^\alpha \right], \tag{20}$$

where $a(\eta)$ is a scalar factor of the universe, $\eta = \int \frac{dt}{a}$ is a conformal time, $\Phi$ and $\Psi$ are gauge-invariant potentials describing scalar metric perturbations. In the homogeneous approximation, the neutrino distribution function is Fermi–Dirac $F_0(q) = \left( e^{q/aT_\nu} + 1 \right)^{-1}$, $q$ is the comoving momentum. Let $f_n^i(q^\alpha, \eta)$ be the linear perturbations of the neutrino $\nu_i$ distribution functions. The Einstein equations for perturbations in Fourier decomposition ($n_\alpha$ is a comoving wave vector) are reduced to

$$-n^2 \Psi = 4\pi G a^2 \left[ \epsilon_m \left( \frac{\delta\epsilon_m}{\epsilon_m} + 3\frac{a'}{a^2}v_m \right) + \epsilon_r \left( \frac{\delta\epsilon_r}{\epsilon_r} + 4\frac{a'}{a^2}v_r \right) \right]$$

$$+ \sum_i \frac{G}{\pi^2 a^2} \int d^3q \left( \sqrt{m_{\nu_i}^2 a^2 + q^2} + \frac{in_\alpha q^\alpha}{n^2} \right) f_n^i(q^\alpha, \eta),$$

$$n^4 (\Psi - \Phi) = \sum_i \frac{G}{\pi^2 a^2} \int d^3q \frac{n^2 q^2 - 3(n_\alpha q^\alpha)^2}{\sqrt{m_{\nu_i}^2 a^2 + q^2}} f_n^i(q^\alpha, \eta), \tag{21}$$

where $' \equiv \frac{d}{d\eta}$, see Pogosyan & Starobinsky (1993) and Pogosyan & Starobinsky (1995b) for details. The first-order corrections to distribution functions $f_n^i(q^\alpha, \eta)$ satisfy the kinetic equations

$$\frac{\partial f_n^i}{\partial \eta} + \frac{in_\alpha q^\alpha}{\sqrt{m_\nu^2 a^2 + q^2}} f_n^i =$$

$$\frac{1}{q} \frac{dF_0}{dq} \left( in_\alpha q^\alpha \sqrt{m_{\nu_i}^2 a^2 + q^2} \Phi - q^2 \Psi \right). \tag{22}$$

Perturbations in the CDM (m) and radiative (r) components obey standard hydrodynamical equations of motion. In terms of the invariant density perturbation $\delta\epsilon$ and velocity $v$, they are:

$$\left( \frac{\delta\epsilon_m}{\epsilon_m} \right)' + \frac{n^2}{a}v_m = 3\Psi', \qquad v_m'/a = \Phi,$$

$$\frac{3}{4} \left( \frac{\delta\epsilon_r}{\epsilon_r} \right)' + \frac{n^2}{a}v_r = 3\Psi', \qquad (v_r/a)' = \frac{1}{4}\frac{\delta\epsilon_r}{\epsilon_r} + \Phi. \tag{23}$$

These equations were solved numerically.

## APPENDIX B: The Emden Equation for the CHDM Halo

From the phase space density constraint, we have

$$n_0 \leq 0.5. \tag{24}$$

Substituting eq. (11) and (24) into (5), we obtain for the neutrinos mass density

$$\rho_{\nu_i}(r) \leq \frac{m_{\nu_i}^4 \sigma^3}{2\pi^2 \hbar^3} \int dy\, y^2 \frac{(1 + \exp(-\mu))}{1 + \exp\left(\frac{E}{\sigma^2} - \mu\right)}. \tag{25}$$

Note, that actual density is smaller by factor 10. The parameter $\mu$ can be fixed by considering the asymptote of the right hand side of eq. (25) at large radii. Indeed, at large $r$, from eq. (25) it follows that $\rho_{\nu_i}(r) \propto \exp(-\phi(r)/\sigma^2)$, similar to the CDM density (10). At large distances their ratio should matches the background ratio $\Omega_{\nu_i} : \Omega_{CDM}$. From this we obtain

$$1 + \exp\mu \approx 10 \cdot \frac{\Omega_{\nu_i}}{\Omega_{CDM}} \cdot \sqrt{\frac{2}{\pi}} \cdot \rho_{CDM}(0) \cdot \left(\frac{m_{\nu_i}^4 \sigma^3}{2\pi^2 \hbar^3}\right)^{-1}. \tag{26}$$

Thus, in our model the integral (25) mildly depends on the CDM central density. Finally, the neutrino density profile can be estimated by formula

$$\rho_{\nu_i}(r) \approx \frac{\Omega_{\nu_i}}{\Omega_{CDM}} \sqrt{\frac{2}{\pi}} \rho_{CDM}(0) \int_0^\infty \frac{dy\, y^2}{\exp\mu + \exp(y^2/2 + \phi(r)/\sigma^2)}, \tag{27}$$

where $\mu$ is given by equation (26).

Poisson's equation for the gravitational potential $\phi(r)$ is

$$\frac{d^2\phi}{dr^2} + \frac{2}{r}\frac{d\phi}{dr} = 4\pi G\left(\rho_{CDM}(r) + \rho_\nu(r)\right), \tag{28}$$

where the CDM and neutrino densities as functions of $\phi(r)$ are given by eqs. (10), (27). It is convenient to introduce the dimensionless variables, for the potential

$$\phi(r) = -u(r)\sigma^2, \tag{29}$$

and for the length

$$r = \frac{z\sigma}{\sqrt{4\pi G \rho_{CDM}(0)}}. \tag{30}$$

Poisson's equation is reduced to the form

$$\frac{d^2u}{dz^2} + \frac{2}{z}\frac{du}{dz} + e^u + \frac{\Omega_{\nu_i}}{\Omega_{CDM}}\sqrt{\frac{2}{\pi}} \int_o^\infty \frac{dy\, y^2}{\exp\mu + \exp(y^2/2 - u(z))} = 0. \tag{31}$$

The boundary conditions at the center $z = 0$ are $u = 0$ and $du/dz = 0$ This is a modification of the standard form of the Emden equation. In the limit of large $r$ the last equation has the solution

$$\rho_{CDM}(r) \approx \frac{\sigma^2}{2\pi G r^2}\left(1 + \frac{\Omega_\nu}{\Omega_{CDM}}\right). \tag{32}$$

Our toy analytic model is based on simple phase space density distributions (9) and (11), which lead to the mass density distributions (10) and (27), similar to the isothermal profiles, see Fig. 3. This allows us to understand an important feature: the central mass density of neutrinos is significantly suppressed by the factor $f$ relatively to the CDM density. Our numerical "coarse-grained" mass density has similar properties, see sec. 4. However, as numerical simulation of sec. 4.2 shows, the realistic cold component density profile (13) is quite different from that of Fig. 3. In principle, the actual central CDM mass density might be very high. There is an interesting question if the central mass density of the hot component is suppressed by factor $f$ but also high? Unfortunately, our analytic model as well as numerical simulation do not address this question. The finding of the phase space density distribution which could correspond to the mass density (13) is much more complicated problem, and numerical study requires much higher resolution. However, we believe that the neutrino central mass density has no cusp and cannot be very high. This is because the neutrinos begin clustering at relatively late stages, well after the clustering of the cold component.

## Figure Captions

**Figure 1:** The transfer function $C(k)$ of the total density fluctuations in the CHDM model with $\Omega_\nu = 0.2$. An upper curve corresponds to the models with only one 4.6 eV massive neutrino; a middle curve – two 2.3 eV neutrinos; and a lower curve – three 1.53 eV neutrinos.

**Figure 2:** Typical density profiles around the peak, solid curve for CDM and dashed curve for 2.3 eV neutrinos. The upper panels corresponds to the top-hat filtering of the initial density field on the scale $10^{13} M_\odot$; the peak height is $\nu = 1.68$. The lower panels corresponds to the filtering on the scale $10^{15} M_\odot$, the peak height $\nu = 2.5$. The left panels corresponds to the redshift $z = 4$; the right – $z = 0$.

**Figure 3:** The left panel shows the ratio of neutrino and CDM density profiles $f(r) = \rho_\nu(r)/\rho_{CDM}(r)$ as function of $r$. The solid curve corresponds to 2.3 eV neutrinos; dotted curve to 4.6 eV neutrinos, and dashed curve to 1.56 eV neutrinos. The right panel shows three pairs of cold (upper curves) and hot (lower curves) particles density profiles for these three models.

**Figure 4:** The left panel shows the density profiles for the cluster (a) with the mass $6.5 h^{-1} 10^{14} M_\odot$; the right panel shows the density profiles for the cluster (b) with mass $5.5 h^{-1} 10^{14} M_\odot$. The upper panels show the ratio of the cold and hot particles densities; normalized in such a way that unity corresponds to the background ratio. The lower panels shows CDM density profiles and parameters of the fitting form (13). The distances are given for $h = 0.5$.

**Figure 5:** The same as in Figure 4 but for the "old" CHDM model with one 7 eV neutrinos.

**Figure 6:** The profile of the velocity dispersions $v_{rms}(r)$ in the clusters. Left panel is for cluster (a), right panel for cluster (b); both for the 2.3 eV model, see Figure 4. The solid curve shows $v_{rms}(r)$ of cold particles; the dashed curves that of the hot particles. The local velocity dispersion, $\sigma_{rms}(r) = <(v - v_f(r))^2>^{1/2}$, is shown as the dash-dotted curves.

**Figure 7:** (a) Observed X-ray surface brightness $\Sigma_X(\theta)$ for the Abell cluster A2256 (Briel & Henry 1994); solid curve shows the $\beta-$model fit (15). (b)The measured X-ray projected temperature $T(r)$ for this cluster. Solid curve is the projected temperature obtained from (17).

**Figure 8:** Solid curve is the total projected surface density $\Sigma(r)_{tot}$ (in units of mean density by Mpc); dashed curve is projected surface density of the cold component $\Sigma(r)_{CDM}$ only.